# POSTER: Using Unit Testing to Detect Sanitization Flaws

Mahmoud Mohammadi, Bill Chu, Heather Richter Lipford
College of Computing and Informatics
University of North Carolina at Charlotte
Charlotte, NC, USA
{mmoham12, billchu, richter}@uncc.edu

## ABSTRACT
Input sanitization mechanisms are widely used to mitigate vulnerabilities to injection attacks such as cross-site scripting. Static analysis tools and techniques commonly used to ensure that applications utilize sanitization functions. Dynamic analysis must be to evaluate the correctness of sanitization functions. The proposed approach is based on unit testing to bring the advantages of both static and dynamic techniques to the development time. Our approach introduces a technique to automatically extract the sanitization functions and then evaluate their effectiveness against attacks using automatically generated attack vectors. The empirical results show that the proposed technique can detect security flaws cannot find by the static analysis tools.

## Categories and Subject Descriptors
D.2.5 [**Software Engineering**]: Testing and debugging – *Code inspections and walk-throughs.* D.4.6 [**Operating Systems**]: Security and Protection – *Verification.*

## 1. INTRODUCTION
Web applications consume data from different inputs. Some of these inputs are originated from untrusted sources, such as user inputs, and referred as untrusted sources. In addition, many of the data from these sources will be used in functions such as sending data to the outputs or accessing the databases referred as sensitive sinks. In the case of injection attacks and more specifically in cross-site scripting attacks, some of these sinks are sensitive to certain characters or keywords, affecting their functionalities and so malicious inputs can change their planned functionalities to dangerous actions. Therefor any arbitrary inputs cannot be used for them as input. Attacks of type command injections or XSS are results of such problems. To deal with this problem web applications use sanitization functions to make the untrusted inputs free of texts that can be interpreted as scripts. But how we can ensure about the correctness of sanitization functions? Although the sanitization process can be applied to all types of the injection attacks we have focused on cross-site scripting (XSS) attacks in this paper. Sanitization to prevent XSS is context-sensitive. Context here means in which place the untrusted source is going to be used after sanitization. There are different contexts as Html, JavaScript, and style sheets and each of them have different sanitization requirements. Meanwhile, based on many practical experiences and successful attacks vectors, there are different sanitization problems[5]: 1) Context inconsistency and 2) Order of the sanitization functions. The conceptual examples shown in Figure 1 demonstrate these problems.

---

*1) <input type='button' onclick=" ...<%= StringEscapeUtils.escapeHtml( UNTRUSTED) %> " />*

*2)<% htmlEsc = StringEscapeUtils.escapeHtml( UNTRUSTED); %>*

*<input type='button' onclick=" ...<%= StringEscapeUtils. escapeJavascriptl ( htmlEsc ) %> " />*

**Figure 1. Context-sensitive sanitization**

---

In the first example (Figure 1) the Html escaping is used in the event (onclick) context, which is a JavaScript context. Html escaping is insufficient to prevent XSS attacks. In the second one, the order of applying the JavaScript escaping function is wrong and it has no effect on the previously sanitized using Html escaping function. Figure 6 and Figure 7 show real attack vectors for such problems for the first and the second problem respectively. The root cause of these problems can be seen better by looking deeper into internal behavior of the browsers. The browsers have different internal interpreters for different grammars such as mentioned before and each of them is sensitive to different characters and keywords (Problem 1). In addition, once these interpreters encounter a keyword and before transferring the control to another interpreter, they may decode some of the input stream characters, causing some issues namely as browser transduction problem [5](Problem 2). Currently, static analysis tools widely used to check whether the web application use any kind of sanitization or not, but these tools can only check the existence of sanitizers and not their correctness. Here correctness means both satisfying the requirements of the target context the sanitizer is designed for and also the order of the sanitizations used in the path from the untrusted sources to security sinks. A single untrusted source could have different sanitization paths (section 2.2) based on different control flows and target security contexts and thus the type and order of sanitizations used in each path should be different.

## 2. APPROACH
Our proposed approach is based on automatic generation of security test cases. Software testing tools and methodologies always has to deal with the structural coverage problem[6]. In the case of evaluating the sanitizations paths spread in different modules of an application, two challenges can be revealed. The first one is finding all sanitization functions applied to untrusted sources across the application modules and the second one is the generating test inputs to maximize the quality and reliability of testing trials. Automatic security test cases generation serve these purposes. This approach is implemented as an IDE plug-in to automatically build security "Test Cases" based on extracted sanitization paths and then evaluating them by injecting attacks scripts. The automated generated test cases composed of 3 different sections (Figure 2): Attack generation, Sanitization path extraction and Attack evaluation. These steps described as below.

```
SanitizationPath-TestCase()
{
atkVec=Attack-Generation();
sanInp=Sanitization-Path(atkVec);
Assert Attack-Evaluation(sanInp);
}
```

**Figure 2. General structure of security test cases**

## 2.1 Attack generation

The goal is to automatically generate attack vectors for the application under unit testing. We want to ensure that, if there is an attack vector capable of exploiting a vulnerable sanitization path, this attack vector can be generated. In other words the false negative rate of this section should be zero or very low. One approach to generate attack vectors is to use attack vector repositories containing different attack patterns such as OWASP XSS evasion list but obviously it is impossible to estimate the false negative rates of such repositories. The other approach is to generate attack scripts based on the specific application context of the injection point. Injection point is the sink or the final destination of the untrusted variable after sanitization. It is the place the attackers try to inject their attack scripts to exploit potential vulnerabilities. As mentioned before, the browsers have different internal contexts which each of them correspond to one grammar. These contexts fall into either an executable context or a non-executable context. The only executable context is defined by the JavaScript grammar. Here the goal of attackers is to trick the browsers to run their attack scripts directly. If the injection point is already in a JavaScript context it could be (at its simplest form) an attack vector to end the current statement and then start a malicious code (attack payload). But if it is in a *non-executable* context such as HTML tag attribute, at first it should change the current context (grammar) to a JavaScript enabled one (e.g. using "*javascript:*" keyword in tag's value) and then run the malicious code. In general an attack vector could have a pattern composed of *pre-escaping characters, attack payload, post-escaping characters*. The Figure 3 shows this pattern in action.

<input type='button' **onclick**=" Func(' **UNTRUSTED**'); " />

| Pre escaping | Attack Payload | Post escaping |
|---|---|---|
| '); | Alert(1); | // |

**Figure 3. Attack vector pattern**

Here attack payload is a character string, which should be a valid statement in the target grammar. In the case of XSS attacks the target grammar is JavaScript. Also pre-escaping and post-escaping strings are completely application and context-sensitive. In other words, these escaping characters will be determined considering the context and the surrounding characters of the injection point. Moreover, considering the flow of data and internal behavior of the browsers in different contexts explained in [5] and also formal published html specifications, we can assume that there is a branching mechanism( e.g. a switch-case statement)  in the browser engine which calls a certain grammar interpreter based on the parsed token and then transfer the control to this interpreter. Using this view, it can be said that an attack vector is an input that tries to modify the source code to prevent the browser interpreting in a planned branch and change control to the JavaScript branch or change the current interpreting flow of characters if the it is already in JavaScript context. We can cast this problem formally based on previous researches in symbolic execution. A path coverage problem is to select a range of input values in such way that a particular point of an application can be reached (after passing constraints from entry points to the destination). The attack script generation problem is a path coverage problem, which aimed to reach a particular branch (JavaScript interpreter) of the application (browser). We define all surrounding characters of the injection point in most recent DOM element as the constraints. If this constraint can be solved, an attack vector exists and if not we are sure that no attack vector exists for this combination of context and constraint. Solution (attack vector) is a string that should be solved using rules of both current and target (if they are different) grammars of injection point. The architecture of the attack generation is shown in  Figure 4.

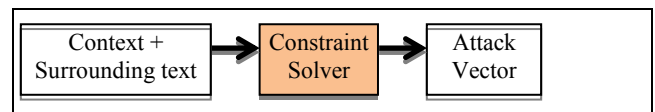

**Figure 4. Attack generation architecture**

 Figure 4 shows attack generation architecture, which contains a constraint solver box customized to solve constraints satisfying different grammars. Obviously solving the constraints depends on their complexities and may be very time consuming[2]. These constraints are all string constraints that should be expressed in regular and context-free grammars [4] in order to be efficiently solved by constraint solver.

## 2.2 Sanitization path extraction

 The goal sanitization path extraction is to build a model of the application that only contains input sanitization logic leaving aside other application specific logic.

```
String login = (String)df.get("username"); //Untrusted Source
login =ESAPI.validator().getValidInput("User", login, ..... );
…
System.err.println(ESAPI.encoder().encodeForHTML(login));
```

**Extracted Sanitization Path:**
*login=ESAPI.validator().getValidInput("User", login, …);*
*ESAPI.encoder().encodeForHTML(login)*

**Figure 5. Sanitization Path Extraction**

A sanitization path, which is called path from now on, is the combination of all sanitization functions applied to one untrusted source and the variables derived from it in the same order appeared in the source code (Figure 5). Two paths are similar if the type and order of their functions are the same. Considering this definition it is very likely that different sources have similar paths but based on the context of their final sensitive operation, they can be vulnerable to different attack scripts. Similar paths mean similar test cases, which required to be merged to one test case. This process uses static analysis techniques such dataflow and control flow analysis to extract the sanitization paths. The important point here is that because untrusted string values can take different control flows before be used in sensitive operations, it is required that all possible control flow structures such as if/else statements and function calls considered. In the proposed approach the developers would declare the sanitization functions to be monitors and extracted. Current limitations are that only server-side functions and only the string type untrusted sources are considered for any analysis.

## 2.3 Attack evaluation

The goal of attack evaluation is to assess whether the extracted sanitization path is vulnerable to the generated attack scripts. There are some challenges for this evaluation. The first one is that because some sanitization flaws, such as browser transduction, can only be revealed when the attack scripts execute in a real browser, thus the attacks should be really executed. This can be (approximately) accomplished using browser components or libraries such as JWebUnit. The second challenge is that some vulnerabilities are triggered only by user interactions as html links or mouse hovers. To deal with this issue, the proposed technique simulates the user interactions using features provided by browser components. It is noteworthy that all of this process is done in a unit-testing framework such as JUnit. Advantages of using unit-testing framework are two folds. The first one is the popularity of these frameworks among the developers, which makes its usage fairly straightforward by efficient utilization of their features such as whole test process automation. The second advantage is early discovery of the vulnerabilities and increase security awareness of developers[7] causing improvements in time and cost of removing security flaws.

## 3. EMPIRICAL RESULTS

We applied the proposed approach to an open source web-based medical application (iTrust) and found a zero-day vulnerability in one of its modules. In this application untrusted input is used in an event context, a JavaScript context, but the sanitization used for this purpose is not matched with the sink's context (Figure 6). In this case the *request.getParameter("forward")* is an untrusted source which is sanitized using *StringEscapeUtils.escapeHtml()* which is not safe for the target context.

---

<input type='button' onclick= "parent.location.href= 'getPatientID.jsp?forward= <%=**StringEscapeUtils.escapeHtml(""** + ( request.getParameter("forward") )) %> ';" … />

Attack vector: *'; alert(1); //*

---

**Figure 6. Sanitization flaw found in the iTrust**

---

String sant= StringEscapeUtils.escapeHtml (source);

sant = StringEscapeUtils.escapejavascript (sant);

tag.innerHTML = '<a onclick="MyFunc(\" + <%= sant %> + '\')">' + sant + '</a>';

**Attack vector:  '); Alert(1);//**

---

**Figure 7. A Conceptual nested sanitization flaw**

Also we applied the proposed technique to a conceptual example containing nested contexts to introduce browser transduction challenge. In this case the source is an untrusted source used in an event context and sanitized for both context of html and JavaScript but the order of sanitization is not correct (Figure 7). The attack script *'); Alert(1);//* at first will be escaped to *");alert(1);//* which will not be changed by the second sanitization because the single quote character escaped as " and because all of the characters are legal, the *"* characters will be decoded to single quote ' at run time by the browser and so causing the attack script to be successful.

## 4. RELATED WORK

Previous research [1] performed heuristic dynamic evaluation of sanitization functions by injecting predefined attack vectors, making it difficult to evaluate false negatives. Our approach generates attack vectors based on the application under testing and can demonstrate low false negatives, given sufficient computing resources Researchers in[3] introduced a vulnerability injector tool(VAIT) for SQL injection. It is not clear this can be generalized to other types of injection attacks. None of these works and many of similar ones have considered unit-testing approach to bring their evaluations into early software development cycle.

## 5. CONCLUSION

We propose a unit testing based approach to detect injection vulnerabilities that can complement static analysis and ensure sanitizations are performed correctly. This approach can be fully integrated into IDEs as a development time plugin, combining static and dynamic security testing features. It means that this integration can be efficiently adjusted to satisfy agile software development life cycle requirements and methodologies.

## 6. ACKNOWLEDGMENT

This research is support in part by NSF grants: 1129190, 1318854.

## 7. REFRENCES

[1] D. Balzarotti, M. Cova, V. Felmetsger, N. Jovanovic, E. Kirda, C. Kruegel, and G. Vigna. Saner: Composing Static and Dynamic Analysis to Validate Sanitization in Web Applications. *In 2008 IEEE Symposium on Security and Privacy (sp 2008)*, pages 387–401. IEEE, May 2008.

[2] J. Fonseca, M. Vieira, and H. Madeira. Evaluation of Web Security Mechanisms Using Vulnerability & Attack Injection. *IEEE Transactions on Dependable and Secure Computing*, 11(5):440–453, Sept. 2014.

[3] J. Thomé, A. Gorla, and A. Zeller. Search-based security testing of web applications. *In Proceedings of the 7th International Workshop on Search-Based Software Testing - SBST 2014*, pages 5–14, New York, New York, USA, 2014.

[4] A. van Deursen, A. Mesbah, and A. Nederlof. Crawl-based analysis of web applications: Prospects and challenges. *Science of Computer Programming*, 97:173–180, Jan. 2015.

[5] J. Weinberger, P. Saxena, D. Akhawe, M. Finifter, R. Shin, and D. Song. A Systematic Analysis of XSS Sanitization in Web Application Frameworks. *Computer Security - ESORICS 2011 SE - 9*, volume 6879 of Lecture Notes in Computer Science, pages 150–171. Springer Berlin Heidelberg, 2011.

[6] X. Xiao, T. Xie, N. Tillmann, and J. de Halleux. Precise identification of problems for structural test generation. In *Proceeding of the 33rd international conference on Software engineering - ICSE '11*, page 611, New York, New York, USA, 2011. ACM Press.

[7] J. Xie, B. Chu, H. R. Lipford, and J. T. Melton. ASIDE: IDE Support for Web Application Security. In *Proceedings of the 27th Annual Computer Security Applications Conference on - ACSAC '11*, page 267,New York, New York, USA, 2011.